\documentstyle[twocolumn,aps,prl,tighten,psfig]{revtex}
\begin{document}
\draft

\def\Tr{\mathop{\rm Tr}}
\def\Var{\mathop{\rm{var}}}
\def\Ei{{\rm Ei}}
\renewcommand{\Im}{\mathop{\rm{Im}}}
\title{
Localization of Light: Dual Symmetry between Absorption and
Amplification}

\author{J.~C.~J.~Paasschens,$^{a,b}$ T.~Sh.~Misirpashaev,$^{a,c}$ and
C.~W.~J.~Beenakker$^a$}
\address{$^a$Instituut-Lorentz, University of Leiden,
                 P.O. Box 9506, 2300 RA Leiden, The Netherlands\\
$^b$Philips Research Laboratories, 5656 AA Eindhoven, The
      Netherlands\\
$^c$Landau Institute for Theoretical Physics, 2 Kosygin Street,
Moscow 117334, Russia}
\date{\today}
\maketitle

\begin{abstract}

We study the propagation of radiation through a disordered waveguide
with
a complex dielectric constant~$\varepsilon$,  and show that dual
systems,
which differ only in the sign of the imaginary part of~$\varepsilon$,
have the same localization
length. Paradoxically, absorption and stimulated emission of
radiation
suppress the transmittance of the waveguide in the same way.

\end{abstract}
\pacs{PACS numbers: 78.45.+h, 42.25.Bs, 72.15.Rn, 78.20.Ci
        {\tt cond-mat/9602048}}
\narrowtext

Localization of electromagnetic waves in a random medium has
attracted much
interest \cite{Joh91}, since the original proposals of John
\cite{Joh84} and
Anderson \cite{And85}. An essential difference with localization of
electrons
is the absence of a conservation law for photons. Light is absorbed
or
amplified---while retaining the phase coherence---if the dielectric
constant  has a non-zero
imaginary part. The intensity of the radiation which has propagated
without
reflection over a distance  $L$ is then multiplied by a factor
$e^{\sigma L}$,
with $\sigma$ negative (positive) for absorption (amplification). The
interplay
of absorption and localization has been studied extensively
\cite{Joh84,And85,Wea93,Yos94,Ram87,Fre94,Gup95}. For the
one-dimensional
problem of
a disordered single-mode waveguide (length $L$, mean free path $l$),
the
result for the transmittance $T$ (being the ratio of transmitted and
incident
flux) is \cite{Ram87,Fre94}:
\begin{equation}
  \langle\ln  T\rangle=(\sigma-l^{-1})\, L,
  \label{eq1}
\end{equation}
where $\langle\cdots\rangle$ denotes an average over disorder.
Eq.~(\ref{eq1}) was derived for $\sigma<0$, corresponding to
absorption.

In this paper we address the question: What happens for
amplification?
Since $T=e^{\sigma L}$ in the absence of reflection for both
positive and negative $\sigma$, one might surmise that
Eq.~(\ref{eq1})
holds both for absorption and amplification. This is correct for
short waveguides.
However, we find that the asymptotic result for $L\to\infty$ is
\begin{equation}
  \langle\ln  T\rangle=(-|\sigma|-l^{-1})\, L+{\cal O}(1).
  \label{1}
\end{equation}
We will show that exponential decay of the transmittance in the case
of
amplification, $\langle\ln  T\rangle\simeq-L/\xi$, is in fact implied
by its exponential decay in the case of absorption, with a duality
relation
between decay lengths:
\begin{equation}
  \xi(\sigma) = \xi(-\sigma).
  \label{xiisxi}
\end{equation}
This duality relation extends beyond the strictly one-dimensional
case, the only essential ingredient being an exponentially decaying
transmittance in an absorbing system.
Contrary to intuition, amplification suppresses the
transmittance in the large-$L$ limit just as much as absorption does.

Experimentally, a random amplifying medium can be realized in a
turbid
laser dye or a powdered laser crystal \cite{Law94,Sha94,Wie95}.
Stimulated emission of radiation leads to a dielectric constant
with a negative imaginary part, corresponding to $\sigma>0$. We
do not present a complete theory for such a ``random laser'',
because we ignore spontaneous emission. (This would correspond
to a source term in the wave equation \cite{Zyu95}, which we do not
include.) Still, because of the different time scales for stimulated
and spontaneous emissions, we believe that a time-resolved experiment
in a waveguide geometry might give evidence for the localization
of stimulated emission, before spontaneous emission sets in.

To prove the duality relation~(\ref{xiisxi}) we consider the
propagation of
monochromatic radiation (scalar amplitude $E$, wavenumber $k$),
described by the Helmholtz equation
\begin{equation}
  {\cal H}E(\vec r)=0,\qquad {\cal H}=\nabla^2+ k^2\varepsilon(\vec
r).
\label{Helm}\end{equation}
(We suppose that all polarization-sensitive phenomena are absent.)
Disorder leads to spatial fluctuations of
the real part $\varepsilon'(\vec r)$ of the dielectric constant.
In the absence of disorder $\varepsilon'=1$.
A non-zero imaginary part $\varepsilon''$ makes the system
non-conservative.
For simplicity we assume a homogeneous $\varepsilon''$.
Its sign determines whether the system is absorbing
($\varepsilon''>0$) or amplifying ($\varepsilon''<0$). The parameter
$\sigma$ introduced
above is related to $\varepsilon''$ by
\begin{equation}
  \sigma = -2k \Im\sqrt{1+i\varepsilon''},
  \label{sigdef}
\end{equation}
where the argument of the square root is chosen in the interval
$(-\pi/2,\pi/2)$. For $|\varepsilon''|\ll1$ one has
$\sigma=-k\varepsilon''$.

\begin{figure}
\psfig{file=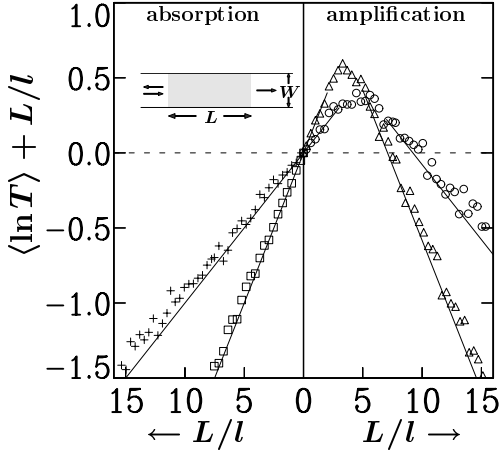%
 ,width=\columnwidth,bbllx=10mm,bblly=30mm,bburx=190mm,bbury=170mm}
\caption{Numerical simulation of the transmittance of a single-mode
waveguide ($W=d$, $k^2=0.5\,d^{-2}$, $\delta\varepsilon=0.2$,
$l=521\,d$,
$N=1$), averaged over $10^4$ realizations of the disorder.
The right half of the figure is for amplification (circles:
$\gamma=0.1$;
triangles: $\gamma=0.2$), the left half is for absorption (crosses:
$\gamma=-0.1
$;
squares: $\gamma=-0.2$). The solid lines are the analytical
asymptotes
from Eq.~(\ref{lTMean}). Their slope is independent of the sign of
$\gamma$,
in agreement with the duality relation~(\ref{xiisxi}). The inset
shows
the geometry considered.}
\label{fig1}
\end{figure}

The dual symmetry underlying Eq.~(\ref{xiisxi}) is formulated in its
general form in terms of scattering matrices.
We assume that the system consists of a scattering region
of length $L$, in which $\varepsilon=\varepsilon'(\vec
r)+i\varepsilon''$,
embedded in an $N$-mode waveguide with $\varepsilon=1$ (see
Fig.~\ref{fig1}, inset).
The scattering matrix
$S$ is a $2N\times2N$ matrix relating incoming and outgoing modes. It
has
the block structure
\begin{equation}
  \quad S={r\ t'\choose t\ r'},
\end{equation}
where $r$, $r'$ are the reflection matrices and $t$, $t'$ the trans-
\noindent
mission matrices. The transmittances and reflectances are defined as
\begin{mathletters}
\label{TRboth}
\begin{eqnarray}
  T=N^{-1}\Tr tt^\dag,&\qquad &R=N^{-1}\Tr rr^\dag,\label{TR}\\
  T'=N^{-1}\Tr t't'{}^\dag,&\qquad &R'=N^{-1}\Tr
r'r'{}^\dag.\label{TRp}
\end{eqnarray}
\end{mathletters}%
Here $T$ and $R$ are the transmitted and reflected flux divided by
the
incident flux from the left. Similarly, $T'$ and $R'$ correspond
to incident flux from the right. By taking the trace in
Eq.~(\ref{TRboth})
we are assuming diffuse illumination, i.e.\ that the incident flux is
equally
distributed over the $N$ modes.
In the absence of gain or loss ($\sigma=0$) the scattering matrix is
unitary,
$SS^\dag=1$. This relation expresses
flux conservation and relies upon Hermiticity of the Helmholtz
operator,
${\cal H}={\cal H}^\dag$ at $\sigma=0$. For non-zero $\sigma$ we have
${\cal H}(\sigma)={\cal H}^\dag(-\sigma)$, which implies the
duality relation
\begin{equation}
  S(\sigma)S^\dag(-\sigma)=1.
\label{GF}\end{equation}

Let us now examine the consequences of the duality relation
(\ref{GF}) for the
reflection and transmission matrices of two systems which differ only
in the
sign of $\sigma$. (We call these systems ``dual''.)
Take $\sigma>0$, so that
$-\sigma$ corresponds to absorption. In the limit $L\to\infty$, all
elements of
$t_{-\sigma}$ and $t'_{-\sigma}$ tend to zero, while $r_{-\sigma}$
and $r'_{-\sigma}$ remain finite.
Expanding the inverse of $S(-\sigma)$ to
first order in the transmission matrices and equating the result to
$S^\dag(\sigma)$, we find
\begin{eqnarray}
  r^\dagger_{\sigma} &=& r^{-1}_{-\sigma}+{\cal O}(t^2),\qquad
  r'^\dagger_{\sigma} = r'^{-1}_{-\sigma}+{\cal O}(t^2),
  \label{rPer}\\
  t^\dagger_{\sigma} &=& -r^{-1}_{-\sigma} t'_{-\sigma}
r'^{-1}_{-\sigma}
  +{\cal O}(t^2).
  \label{tPer}
\end{eqnarray}
We introduce the  transmission and reflection eigenvalues
${\cal T}^{\vphantom1}_n$,
${\cal T}'_n$, ${\cal R}^{\vphantom1}_n$, ${\cal R}'_n$, being the
eigenvalues of, respectively, ${\mathbf T} = tt^\dagger$,
${\mathbf T'} = t't'^\dagger$,
${\mathbf R} = rr^\dagger$,
${\mathbf R'} = r'r'^\dagger$.
Because of time-reversal symmetry $S(\sigma)S^*(-\sigma)=1$.
Together with Eq.~(\ref{GF}) this implies that $S$ is a symmetric
matrix. It follows that $t'=t^{\rm T}$, hence
${\cal T}^{\vphantom1}_n = {\cal T}'_n$ and
$T=T'$.
The reflectances $R$ and $R'$ may differ.
Eq.~(\ref{rPer}) directly
yields a duality relation for the reflection eigenvalues in the limit
$L\to\infty$,
\begin{equation}
  {\cal R}^{\vphantom1}_n(\sigma) = {\cal R}_n^{-1}(-\sigma).
\end{equation}
Eqs.~(\ref{rPer}) and (\ref{tPer}) together imply that the matrices
${\mathbf T}'_{-\sigma}{\mathbf R}_{-\sigma}^{-1}$ and
${\mathbf T}^{\vphantom1}_{\sigma}{\mathbf R}_{\sigma}'^{-1}$
have the same eigenvalues.
The duality relation for the transmission eigenvalues follows from
the
following lemma:

Let $A(L)$ be a matrix function of $L$
with exponentially decreasing eigenvalues $a_n(L)$.
The eigenvalue localization lengths $\xi_n$ are defined by
$\xi_n^{-1} =-\lim_{L\to\infty} L^{-1}\ln a_n(L)$.
Let $B(L)$ be another non-singular matrix function whose elements
remain finite as $L\to\infty$.
Then the matrix $AB$ has the same eigenvalue localization lengths as
$A$.

It follows that the matrices
${\mathbf T}^{\vphantom1}_{\sigma}$,
${\mathbf T}_{\sigma}^{\vphantom1}{\mathbf R}_{\sigma}'^{-1}$,
${\mathbf T}_{-\sigma}'{\mathbf R}_{-\sigma}^{-1}$,
${\mathbf T}'_{-\sigma}$,
and hence
${\mathbf T}_{-\sigma}$ all
have the same eigenvalue localization lengths. Explicitly,
\begin{equation}
  -\lim_{L\to\infty}L^{-1}\ln {\cal T}_n(\sigma)
  =-\lim_{L\to\infty}L^{-1}\ln {\cal T}_n(-\sigma).
\end{equation}
The transmittance $T=N^{-1}\sum_n{\cal T}_n$
is dominated by the largest transmission eigenvalue, which is the
${\cal
T}_n$ with the largest localization length: $\xi =
\max(\xi_1,\xi_2,\ldots,\xi_N)$.
This completes the proof of Eq.~(\ref{xiisxi}), since we have shown
that
all, and in particular the largest, transmission eigenvalues of dual
systems have the same localization length.

The case $N=1$ of a single-mode waveguide can be analyzed in
more detail. We assume that the wavelength $\lambda$ is much
smaller than both $l$ and $1/|\sigma|$. (This is not a restrictive
assumption for an optical system.)
The joint probability distribution $P(R,T,L)$ of reflectance and
transmittance evolves with increasing $L$ according to a
Fokker-Planck equation,
\begin{eqnarray}
   l\frac{\partial P}{\partial L} =
  &-&\frac{\partial}{\partial R} [(1-R)^2 + 2\gamma R]P
  +\frac{\partial^2}{\partial R^2} R(1-R)^2P\nonumber\\
  &-&\frac{\partial  }{\partial T} T(\gamma-1+R)P
  +\frac{\partial^2}{\partial T^2} T^2RP\nonumber\\
  &-&2\frac{\partial^2}{\partial R\,\partial T}TR(1-R)P,
  \label{FPETR}
\end{eqnarray}
where we have abbreviated $\gamma = \sigma l$. For $\gamma<0$
(absorption),
this equation is equivalent to the moment equation of Freilikher,
Pustilnik,
and Yurkevich \cite{Fre94}. For $\gamma>0$ (amplification) their
method
of moments cannot be used, because all moments of $R$ diverge as
$L\to\infty$.
The derivation of Eq.~(\ref{FPETR}) proceeds along the lines of
Ref.~\cite{Mel88b}, where the case $\gamma=0$ was considered. On
integration
over $T$ it reduces to a well known \cite{Ger59,Koh76,Pra94}
Fokker-Planck equation for $P(R,L)=\int dT\,P(R,T,L)$.
The limit $L\to\infty$ of $P(R,L)$ was studied in
Refs.~\cite{Koh76,Pra94,Bee95}.
In terms of the variable $\mu=1/(R{-}1)$ it reads
\begin{equation}
  P(\mu)=\cases{
  2\gamma\,e^{-2\gamma\mu}\,\theta(\mu)&  for $\gamma>0$,\cr
  -2\gamma\,e^{-2\gamma(1+\mu)}\,\theta(-1-\mu)&  for $\gamma<0$,}
  \label{DistmuStat}
\end{equation}
where $\theta(x)=1$ for $x>0$ and 0 otherwise.
Using this asymptotic distribution we have computed from
Eq.~(\ref{FPETR})
the first two moments of $\ln T$ in the large-$L$ limit. The result
for
the average is
\begin{mathletters}\label{lTMean}
\begin{eqnarray}
  \langle\ln  T\rangle&=&-(1+|\gamma|)\,L/l+2c(\gamma),\label{lT}\\
  c(\gamma)&=&\cases{
    0                                                     & for
$\gamma<0$, \cr
      {\bf C}+\ln  2\gamma-e^{2\gamma}\Ei(-2\gamma) & for
$\gamma>0$,}
  \label{lTint}
\end{eqnarray}\end{mathletters}%
where $\bf C$ is Euler's constant and
$\Ei(x)=\int_{-\infty}^xdt\,e^t/t$ is the exponential integral
($c(\gamma)\approx -2\gamma\ln\gamma$ if $0<\gamma\ll1$).
For $\gamma<0$, Eq.~(\ref{lTMean}) reduces to the result
of Refs.~\cite{Ram87,Fre94}. The result for $\gamma>0$ is new, and
demonstrates
that the inverse localization length
$\xi^{-1}(\sigma)=(1+|\gamma|)\,l^{-1}=
l^{-1}+|\sigma|$ is indeed independent of the sign of $\sigma$---in
accordance
with the general duality relation~(\ref{xiisxi}). The result for the
variance is
\begin{equation}
  \Var\ln  T= 2\Bigl[1+2|\gamma| e^{2|\gamma|}{\rm
Ei}(-2|\gamma|)\Bigr]
  \, L/l +{\cal O}(1),
  \label{lTVar}
\end{equation}
in agreement with Ref.~\cite{Fre94} for $\gamma<0$.
Note that $\sqrt{\Var\ln  T}\ll\langle\ln  T\rangle$ for $L/l\gg1$.
Evaluation of higher moments shows that the distribution
of $\ln  T$ tends to a Gaussian for $L\to\infty$.
(The tails are non-Gaussian, but contain negligible weight.)

These results hold in the large-$L$ limit.
For short waveguides instead of Eq.~(\ref{lTMean}) one has
$\langle \ln T\rangle = -(1-\gamma)\, L/l$. The crossover length can
be
estimated as $L_c\simeq l\,c(\gamma)/|\gamma|$.
Below this length
stimulated emission {\em enhances} transmission through the
waveguide.
On larger length scales stimulated emission {\em reduces}
transmission.
In contrast, the reflectance is enhanced on
every length scale\cite{Koh76,Pra94,Bee95}.

\begin{figure}
\psfig{file=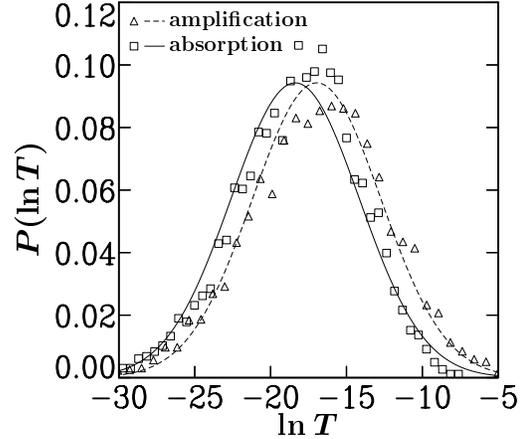%
 ,width=\columnwidth,bbllx=10mm,bblly=30mm,bburx=190mm,bbury=170mm}
\caption{Probability distribution of the logarithm of the
transmittance
of a single-mode waveguide, for
$L/l=15.4$ and $\gamma = 0.2$ (triangles, dashed curve),
$\gamma=-0.2$ (squares, solid curve). The data points are provided by
a numerical simulation (same parameters as in
Fig.~\protect\ref{fig1}),
the curves are a Gaussian distribution of $\ln T$ with mean and
variance
given by Eqs.~(\protect\ref{lTMean}) and~(\protect\ref{lTVar}). There
is
a
slight offset
between the distributions for absorption and amplification because
the system is not fully in the large-$L$ limit.}
\label{fig2}
\end{figure}

\begin{figure}
\psfig{file=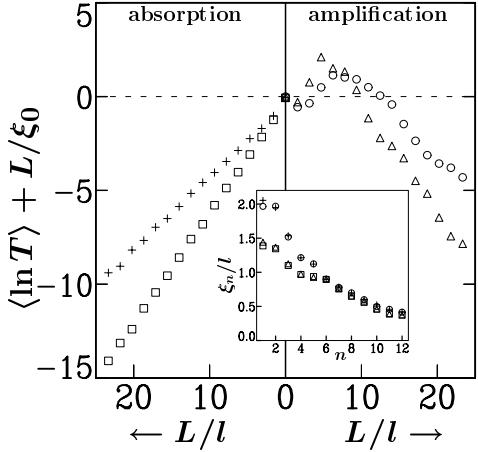%
 ,width=\columnwidth,bbllx=10mm,bblly=30mm,bburx=190mm,bbury=170mm}
\caption{Numerical simulation of the transmittance of a multi-mode
waveguide ($W=25\,d$, $k^2=2.0\,d^{-2}$, $\delta\varepsilon=0.375$,
$l=29.6\,d$, $N=12$), averaged over 50 realizations of the disorder.
The parameter $\xi_0=(N+1)l/2$ is the localization length of the
system
in the absence of absorption or amplification.
The right half of the figure is for amplification
(circles: $\sigma=0.0035\,d^{-1}$; triangles:
$\sigma=0.0071\,d^{-1}$),
the left half is for absorption
(crosses: $\sigma=-0.0035\,d^{-1}$; squares:
$\sigma=-0.0071\,d^{-1}$).
The inset shows the eigenvalue localization lengths,
$\xi_n^{-1}\equiv-\lim_{L\to\infty}L^{-1}\ln{\cal T}_n$.
These lengths $\xi_n$ were computed from
the $L$-dependence of $T_n$ for $L$ up to $40\,l$ and a single
realization
of the disorder. The duality between
absorption and amplification is verified with good accuracy.}
\label{fig3}
\end{figure}

To test these analytical predictions for $N=1$, and to investigate
also
the multi-mode case, we have numerically solved a discretized
version of the Helmholtz equation~(\ref{Helm}), on a two-dimensional
square lattice (lattice constant $d$, length $L$, width $W$).
The real part $\varepsilon'$ of the dielectric constant was chosen
randomly from site to site with a uniform distribution between
$1\pm\delta\varepsilon$. The scattering matrix
for the multi-mode case was computed using
the recursive Green's function technique, originally developed for
the electronic Anderson model \cite{Bar91}. (For the case $N=1$
a transfer-matrix method \cite{Gup95} turned out more convenient.)
Simulations with $\varepsilon''=0$ were used to obtain $l$,
from the relation\cite{Dor84}
\begin{equation}
  -\lim_{L\to\infty} L^{-1} \langle\ln T\rangle =
  [\case12(N+1)l]^{-1}.
  \label{xinuldef}
\end{equation}
The parameter $\sigma$ was determined from Eq.~(\ref{sigdef}).
Results for the single-mode case are shown in Figs.~\ref{fig1}
and~\ref{fig2}, and for the multi-mode case in Fig.~\ref{fig3}.
The duality relation between the localization lengths for absorption
and amplification is verified with good accuracy, both for the
single-
and for the multi-mode case. Furthermore, for $N=1$ we find good
agreement with the results (\ref{lTMean})--(\ref{lTVar}) of the
Fokker-Planck equation.

In conclusion, we have demonstrated that stimulated emission of
radiation in a disordered waveguide {\em reduces} the decay
length, in the same way as absorption does.
This paradoxical result is an immediate consequence of the exact
duality
relation (\ref{GF}) between the scattering matrices of two systems
with complex conjugated dielectric constants. The dual symmetry
between
absorption and amplification has been supported by an explicit
computation of the decay lengths, both analytically (for the
single-mode case) and numerically (for the single- and multi-mode
cases).

We acknowledge useful discussions with P.~W.~Brouwer and K.~M.~Frahm.
This work was supported by the Dutch Science Foundation NWO/FOM.

Note added: Numerical results for the single-mode case supporting
Eq.~(\ref{1}) have been published by Z.Q. Zhang, Phys. Rev. B {\bf 52},
7960 (1995).

\end{document}